\documentclass[aps,prl,reprint,groupedaddress]{revtex4-2}

\usepackage{amsmath}
\usepackage{graphicx}
\usepackage{siunitx}

\begin{document}


\title{Collisionless Shocks Mediated by Shear-Flow Magnetic Fields in Ultraintense-Laser-Produced Counter-Streaming Plasmas}



\author{Jun-Yi Lu}
\affiliation{Zhejiang Institute of Modern Physics, Institute of Astronomy, School of Physics, Zhejiang University, Hangzhou 310027, China}

\author{Kai Wang}
\affiliation{Zhejiang Institute of Modern Physics, Institute of Astronomy, School of Physics, Zhejiang University, Hangzhou 310027, China}

\author{Jin-Long Jiao}
\email[]{jiao.jl@zju.edu.cn}
\affiliation{Zhejiang Institute of Modern Physics, Institute of Astronomy, School of Physics, Zhejiang University, Hangzhou 310027, China}


\date{\today}

\begin{abstract}
The formation of ion-Weibel-mediated collisionless shocks (IW-CSs) in ultraintense-laser-produced counter-streaming plasmas is investigated using particle-in-cell simulations. Analysis of the underlying microphysics reveals that a shear-flow ion-Weibel instability generates magnetic fields, which isotropize the incoming flow and mediate shock formation. An analytical expression for the shear-flow magnetic field is derived, and a scaling law relating the magnetic field amplitude to the laser intensity is established. This mechanism reduces the shock formation time and the required laser energy by three and two orders of magnitude, respectively, compared to high-power laser experiments, making it feasible to produce IW-CSs using existing multi-kilojoule, picosecond ultraintense laser facilities.
\end{abstract}


\maketitle

Collisionless shocks are ubiquitous in a wide range of astrophysical environments, such as active galaxy nuclei, pulsar wind nebulae, and supernovae remnants \cite{Blandford,Piran,Waxman}. These shocks contain turbulent magnetic fields, which can produce high-energy cosmic rays with a power-law energy spectrum through the first-order Fermi acceleration mechanism \cite{Jones,Fermi,Branch, Baade,Koyama}. Observable collisionless shocks in the universe are rare, which hinders in-depth studies of their formation, evolution, and role in cosmic ray acceleration. Such shocks have recently been observed in laser-plasma experiments relevant to laboratory astrophysics. These experiments serve as a valuable complement to traditional astronomical observations by reproducing key astrophysical shock phenomena via scaling transformations \cite{Remington, Ryutov}.

Ion-Weibel-mediated collisionless shocks (IW-CSs) are of interest in laboratory astrophysics due to their potential role in generating high-energy cosmic rays \cite{Kato, Spitkovsky, Spitkovsky2, Kato2, Grassi}. Their formation does not require an external magnetic field, relying solely on the self-generated magnetic fields produced by the ion-Weibel instability (IWI). The IWI, a kinetic micro-instability driven by anisotropy in the ion velocity distribution, is readily excited in high-speed counter-streaming plasma systems \cite{Weibel, Fried, Ross, Fox, Yuan, Huntington, Manuel}. Through self-generated magnetic fields, it deflects ions and converts the kinetic energy of the plasma flow into internal energy, thereby forming collisionless shocks \cite{Moiseev, Medvedev}. Generating IW-CSs in the laboratory remains challenging. The most substantial experimental evidence to date was reported by Fiuza $et~al.$, who observed this phenomenon in counter-streaming plasmas produced by high-power laser beams at the National Ignition Facility (NIF) \cite{Fiuza}. The experiment required a total laser energy of nearly \SI{1}{MJ}, as shock formation demands large characteristic spatial (centimeter-scale) and temporal (tens of nanoseconds) scales \cite{Park, Ross2, Fiuza}. Currently, NIF is the only facility capable of delivering such energy. Although upcoming facilities such as the Laser Megajoule (LMJ) and Shenguang-IV (SG-IV) are designed for megajoule-level output, they are not yet operational for such experiments. This limitation restricts experimental progress on IW-CSs.

Ultraintense lasers (intensity $>\SI{E18}{W/cm^2}$), which exceed conventional high-power lasers by several orders of magnitude, can generate dense, ultrafast plasma flows. This reduces the spatial and temporal scales needed for shock formation and may lower the required laser energy. Fiuza $et~al.$ were among the first to propose a scheme using such lasers to produce collisionless shocks \cite{Fiuza2, Ruyer}. In their setup, a linearly polarized ultraintense laser interacts with a solid target to form a shock. However, strong laser-driven electron heating leads to mediation by the electron-Weibel instability rather than the IWI. Although the resulting ion deflection resembles that in IW-CSs, the underlying mechanisms are fundamentally different. To suppress electron heating and allow full development of the IWI, Grassi $et~al.$ proposed using an obliquely incident S-polarized ultraintense laser \cite{Grassi2}. Yet this approach faces experimental challenges: laser pre-pulses, difficult to fully eliminate, typically create a pre-plasma on the target. In such a pre-plasma, even S-polarized light can cause substantial electron heating, preventing the desired conditions from being achieved. More recently, Kuramitsu $et~al.$ simulated colliding plasma flows driven by two ultraintense lasers \cite{Kuramitsu}. Using relatively low total energy ($\sim \SI{10}{J}$), their simulations observed IWI formation but not an IW-CS.

In this Letter, we demonstrate the generation of IW-CSs in counter-streaming plasmas driven by two picosecond-scale ultraintense lasers. Using relativistic PIC simulations, we examine the underlying shock formation physics. Our results reveal an IWI characterized by shear-flow velocity distributions in ultraintense-laser-produced plasmas. This instability generates spatially confined, high-amplitude magnetic fields that mediate shock formation within picoseconds, nearly three orders of magnitude faster than in high-power laser experiments. These shear-flow magnetic fields reduce the required laser energy, enabling IW-CSs production using existing multi-kilojoule, picosecond-scale ultraintense laser facilities.

We employ two-dimensional particle-in-cell simulations using the EPOCH code \cite{Arber} to study the formation of IW-CSs in counter-streaming plasmas. These plasmas are generated through the interaction of ultraintense laser pulses with preionized unmagnetized electron-proton plasmas. The laser-plasma interaction produces hot electrons that propagate to the target rear and drive high-temperature, high-velocity plasma flows via target normal sheath acceleration (TNSA) mechanism. The simulation domain extends from \SI{-500}{\mu m} to \SI{500}{\mu m} in the $x$-direction and \SI{-150}{\mu m} to \SI{150}{\mu m} in the $y$-direction. Two identical P-polarized laser pulses with a wavelength of \SI{1}{\mu m} enter from the left and right boundaries along the $x$-axis. The laser pulse has a temporal profile consisting of a \SI{0.1}{ps} linear up-ramp (from zero) followed by a \SI{4}{ps} flat-top region at the peak intensity. For the different cases, the peak intensity ranges from $3.45\times10^{19}$ to \SI{5.52E20}{W/cm^2}, corresponding to normalized amplitudes of $a_0=5-20$. Two symmetric targets are initialized and are uniform along $y$-direction. For the left target, the pre-plasma extends exponentially from $x=\SI{-450}{\mu m}$ to \SI{-400}{\mu m} with a \SI{7}{\mu m} scale length, reaching 25$n_c$ at $x=\SI{-400}{\mu m}$, where it connects to a uniform bulk plasma (25$n_c$) extending to $x=\SI{-150}{\mu m}$. The reduced bulk density (25$n_c$ is well below the density of a realistic solid target) is used to avoid numerical heating, but results in higher ion velocities from hole-boring acceleration at the target front. To suppress the artificially high velocities, ions in the region $x<\SI{-350}{\mu m}$ are fixed. Initial electron and ion temperatures are \SI{1}{keV} and \SI{0.5}{keV}, respectively. The grid resolution is 1/12\SI{}{\mu m} (totaling $4.32\times10^7$ cells). Each cell contains 50 macro-electrons and 50 macro-ions initially. Open boundaries are applied in $x$-direction, with periodic conditions in $y$-direction. The real ion-to-electron mass ratio (1836) is used to avoid distortion in shock formation \cite{Ryutov, Ruyer2}.

\begin{figure}
\includegraphics[width=8cm]{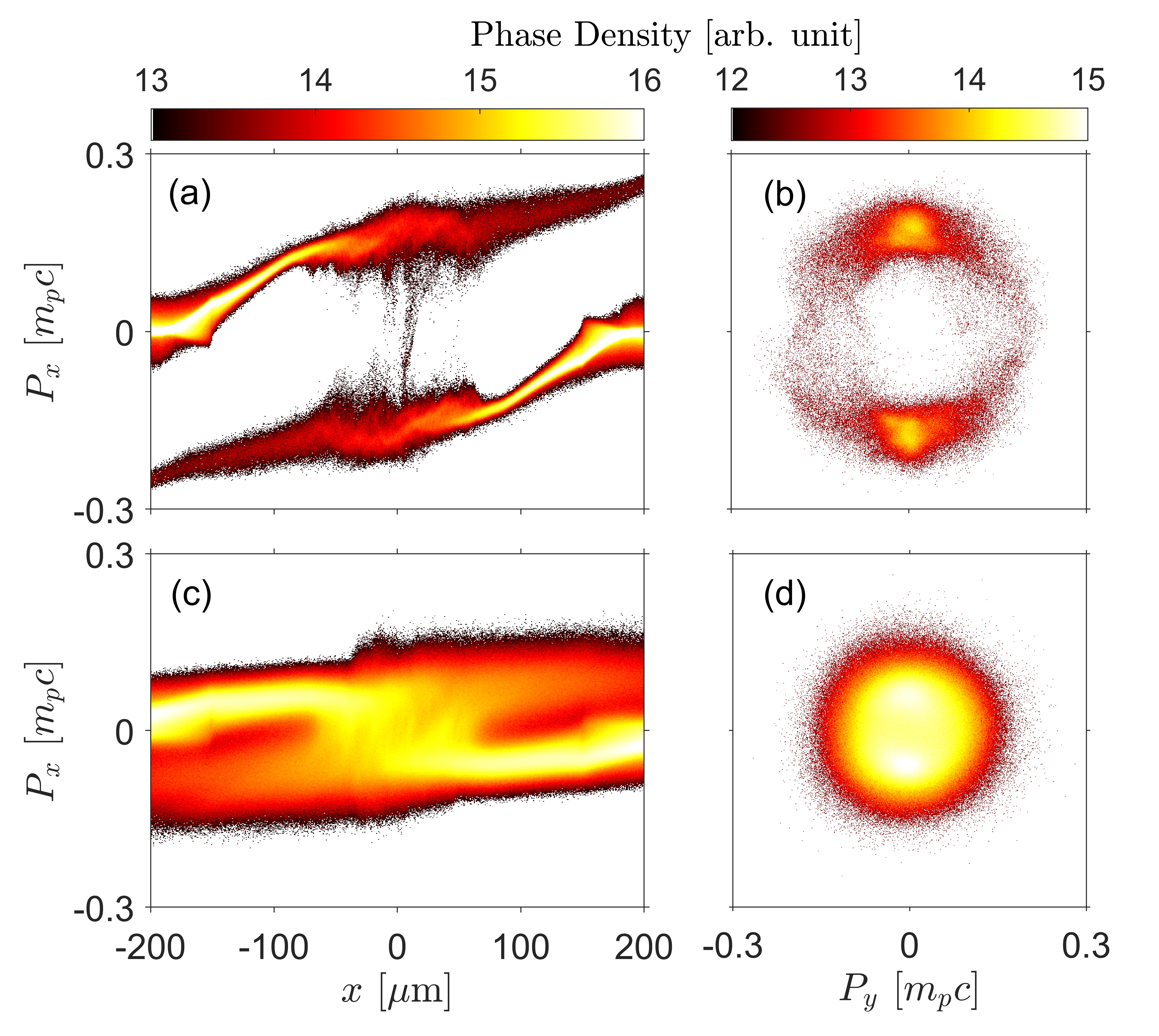}
\caption{Ion phase-space density distributions (logarithmic scale) in counter-streaming plasmas produced by ultraintense laser ($a_0=10$). (a,c) $x-P_x$ distribution. (b,d) $P_x-P_y$ distribution for ions between $x=\SI{-20}{\mu m}$ and \SI{20}{\mu m}. (a,b) and (c,d) show simulation results at \SI{6}{ps} and \SI{16}{ps}, respectively.}
\end{figure}

Ion phase-space signatures confirm the formation of a collisionless shock in ultraintense-laser-produced counter-streaming plasmas. Collisionless shock is a dissipative structure whose formation can be identified by downstream ion thermalization. Fig.\ 1 shows ion phase-space distributions in counter-streaming plasmas. At the early stage (\SI{6}{ps}), the two plasma flows largely interpenetrate with only minor ion thermalization (Figs.\ 1a-b), indicating no shock has formed. After a period of evolution (\SI{16}{ps}), the two plasma flows stagnate and thermalize in the central region (Figs.\ 1c-d), confirming the shock formation. The collisionless nature of the shock is verified by estimating the ion mean free path using $\lambda_{ii}=m_p^2u_0^4/4\pi e^4n_e\mathrm{ln}\Lambda$ \cite{Spitzer}. With $u_0\sim0.1c$, $n_e\sim10n_c$, and $\mathrm{ln}\Lambda \sim10$, the resulting $\lambda_{ii}\sim \SI{30}{m}$ far exceeds the shock scale.

\begin{figure}
\includegraphics[width=8cm]{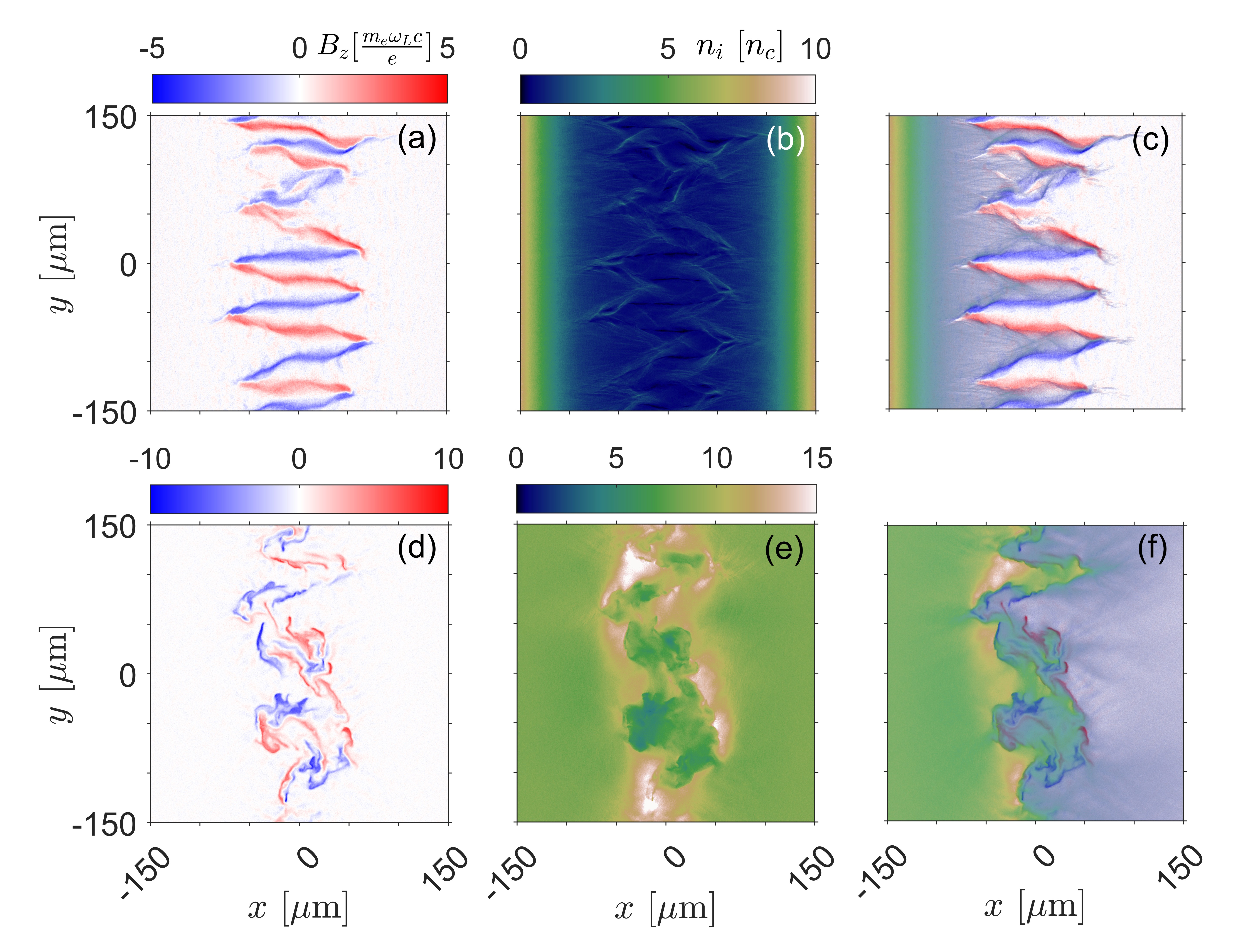}
\caption{Self-generated magnetic field and ion density distributions in counter-streaming plasmas produced by ultraintense laser ($a_0=10$). (a,d) Self-generated magnetic field. (b,e) Ion density. (c,f) Magnetic field overlaid with ion density of left plasma flow. (a-c) and (d-f) show simulation results at \SI{6}{ps} and \SI{16}{ps}, respectively.}
\end{figure}

The observed collisionless shock is mediated by shear-flow magnetic fields. At the early stage (\SI{6}{ps}), self-generated magnetic fields and ion density exhibit quasi-periodic filamentary structures (with an average filament width of $\sim 3c/\omega_{pi}$) consistent with the IWI characteristics (Figs.\ 2a-b). However, unlike classical IWI, the magnetic fields localize specifically within shear layers rather than throughout the entire filament (Fig.\ 2c). Remarkably, the magnetic field remains localized at the shear interface throughout its evolution, even as the overall structure becomes disordered (Figs.\ 2d-f). The shear-flow magnetic fields, though narrow in width, exhibit sufficient strength to deflect ions. Simulations reveal an ion gyroradius of $r_c=m_iu_0c/Bq\approx \SI{15}{\mu m}$ ($B\sim \SI{200}{MG}$ and $u_0\sim0.1c$), closely matching the width of shear-flow magnetic field ($\sim \SI{10}{\mu m}$). This quantitative agreement demonstrates that the shear-flow magnetic field can effectively thermalize ions, thereby enabling shock formation. In addition, electrostatic diagnostics revealed no characteristic signatures of electrostatic collisionless shocks (a localized strong field and periodic oscillations \cite{Tidman, Jiao}). This rules out an electrostatic instability as the mediating mechanism.

\begin{figure}
\includegraphics[width=8cm]{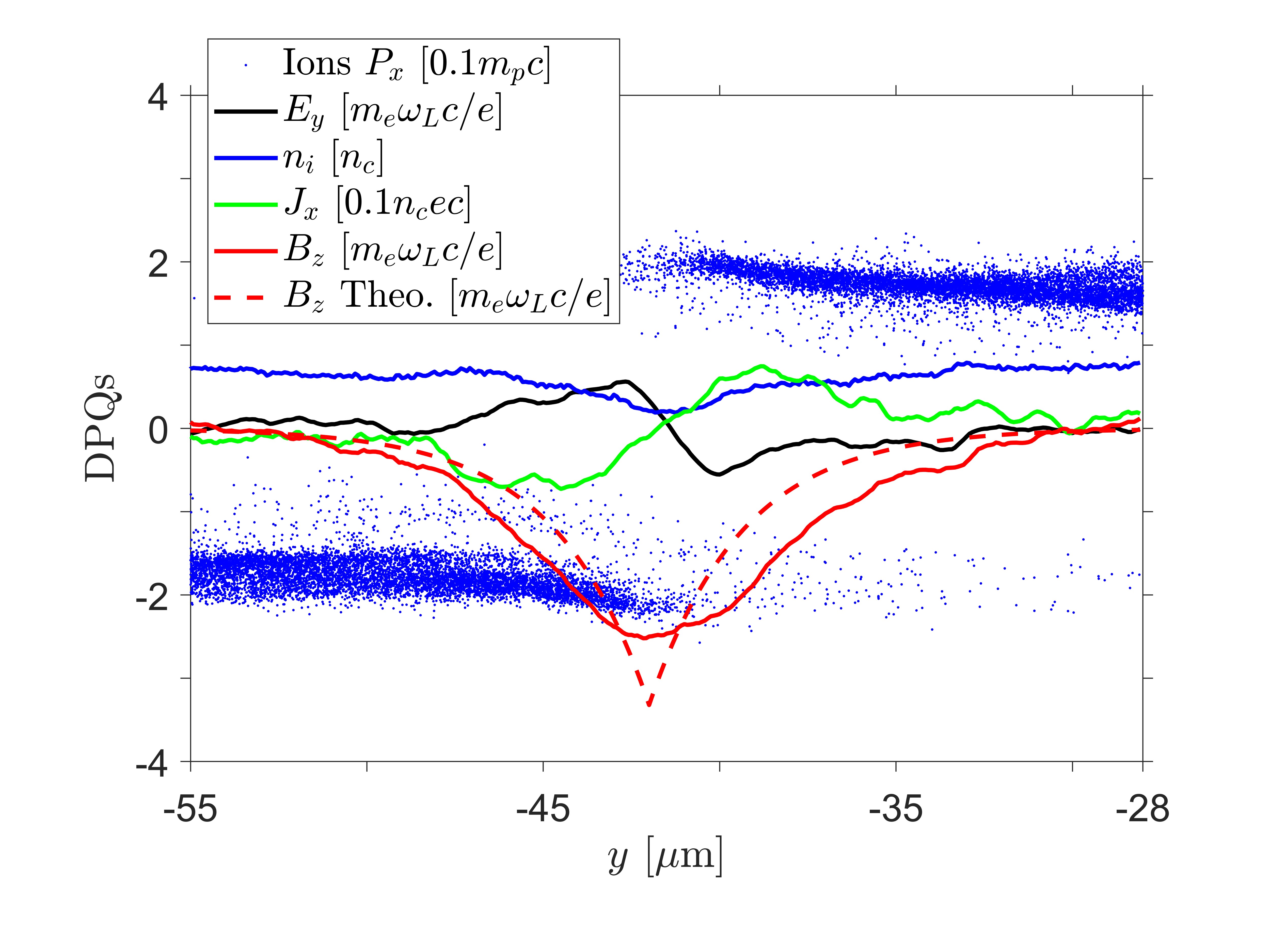}
\caption{Profiles of dimensionless physical quantities (DPQs) along the $y$-direction in a local region ($x=\SI{-20}{\mu m}$ to \SI{15}{\mu m}, $y=\SI{-55}{\mu m}$ to \SI{-28}{\mu m}) at simulation time $t=\SI{6}{ps}$.}
\end{figure}

Shear-flow magnetic fields represent a specialized form of the IWI magnetic fields. The IWI magnetic fields originate from net current densities generated by non-uniform ion fluid velocities \cite{Spitkovsky, Ruyer2}. In classical IWI scenarios, ion fluid velocities exhibit continuous filamentary structures. In contrast, the shear-flow magnetic field arises from a discontinuous ion fluid velocity \cite{Jiao2}, as illustrated in Fig.\ 3 (blue dots). Electrons on either side of the shear interface undergo rapid mixing via the kinetic Kelvin-Helmholtz instability, creating dipolar net current densities at the shear interface (green solid line). This subsequently generates unipolar shear-flow magnetic fields (red solid line) \cite{Alves, Grismayer}. Under the influence of the shear-flow magnetic field, ions on either side of the shear interface tend to separate spatially, resulting in a density gradient (blue solid line). Such density gradients induce electrostatic fields according to the generalized Ohm's law (black solid line). The balance between Lorentz forces and electrostatic forces allows for the stable maintenance of the shear-flow structure.

The one-dimensional expression for shear-flow magnetic fields can be derived through solving the ion current screening (ICS) equation \cite{Jiao2}. For relativistic electron temperatures, we define the average relativistic electron mass as $\bar{\gamma}m_e$, resulting in the ICS equation in CGS units,
\begin{equation}
\frac{c}{4\pi en_e}\nabla^2\textbf{B}-\frac{e}{\eta\bar{\gamma}m_ec}\textbf{B}+\nabla\times\textbf{u}_i=0,
\end{equation}
where $\textbf{u}_i$ is ion fluid velocity and $\eta$ is a kinetic correction factor determined from kinetic simulations. For one-dimensional case, assuming that the discontinuity of shear-flow is at $y=0$. When $y<0$, the ion fluid velocity is $\textbf{u}_i=-u_0\hat{\textbf{e}}_x$ (along $x$-direction), when $y>0$, $\textbf{u}_i=u_0\hat{\textbf{e}}_x$. For all positions of $y\neq0$, Eq.\ 1 is simplified as,
\begin{equation}
\frac{c}{4\pi en_e}\frac{\partial^2B_z}{\partial y^2}-\frac{eB_z}{\eta\bar{\gamma}m_ec}=0.
\end{equation}
Eq.\ 2 only need to be solved over interval $y\in(-\infty,0)$ due to the even symmetry of the shear-flow magnetic field. The general solution of Eq.\ 2 is $B_z=c_1e^{y/\sqrt{\eta\bar{\gamma}}l_s}+c_2e^{-y/\sqrt{\eta\bar{\gamma}}l_s}$, where $l_s=\sqrt{m_ec^2/4\pi n_ee^2}$ is the skin depth of plasma. $c_1$ and $c_2$ are undetermined coefficients, which can be determined by two boundary conditions. The first boundary condition is that as $y\to-\infty$, $B_z=0$, so $c_2=0$. The second boundary condition is that as $y\to0$, electron fluid velocity $u_e=0$, here is $\partial B_z/\partial y=-4\pi n_eeu_0/c$ according to the Ampere's law, so $c_1=-\sqrt{\eta\bar{\gamma}}m_e\omega_{pe}u_0/e$ ($\omega_{pe}=\sqrt{4\pi n_ee^2/m_e}$ is the electron plasma frequency). Finally, the shear-flow magnetic field is
\begin{equation}
B_z=-\sqrt{\eta\bar{\gamma}}\frac{m_e\omega_{pe}u_0}{e}e^{y/\sqrt{\eta\bar{\gamma}}l_s}.
\end{equation}
Eq.\ 3 (red dashed line) agrees with the simulation results (red solid line) as illustrated in Fig.\ 3. The parameters $\bar{\gamma}\approx25$ and $\eta=10$ are determined from the simulation data. The simulation results exhibit lower amplitudes and broader profiles compared to the predictions of Eq.\ 3, which can be attributed to minor ion mixing at the shear interface.

\begin{figure}
\includegraphics[width=8cm]{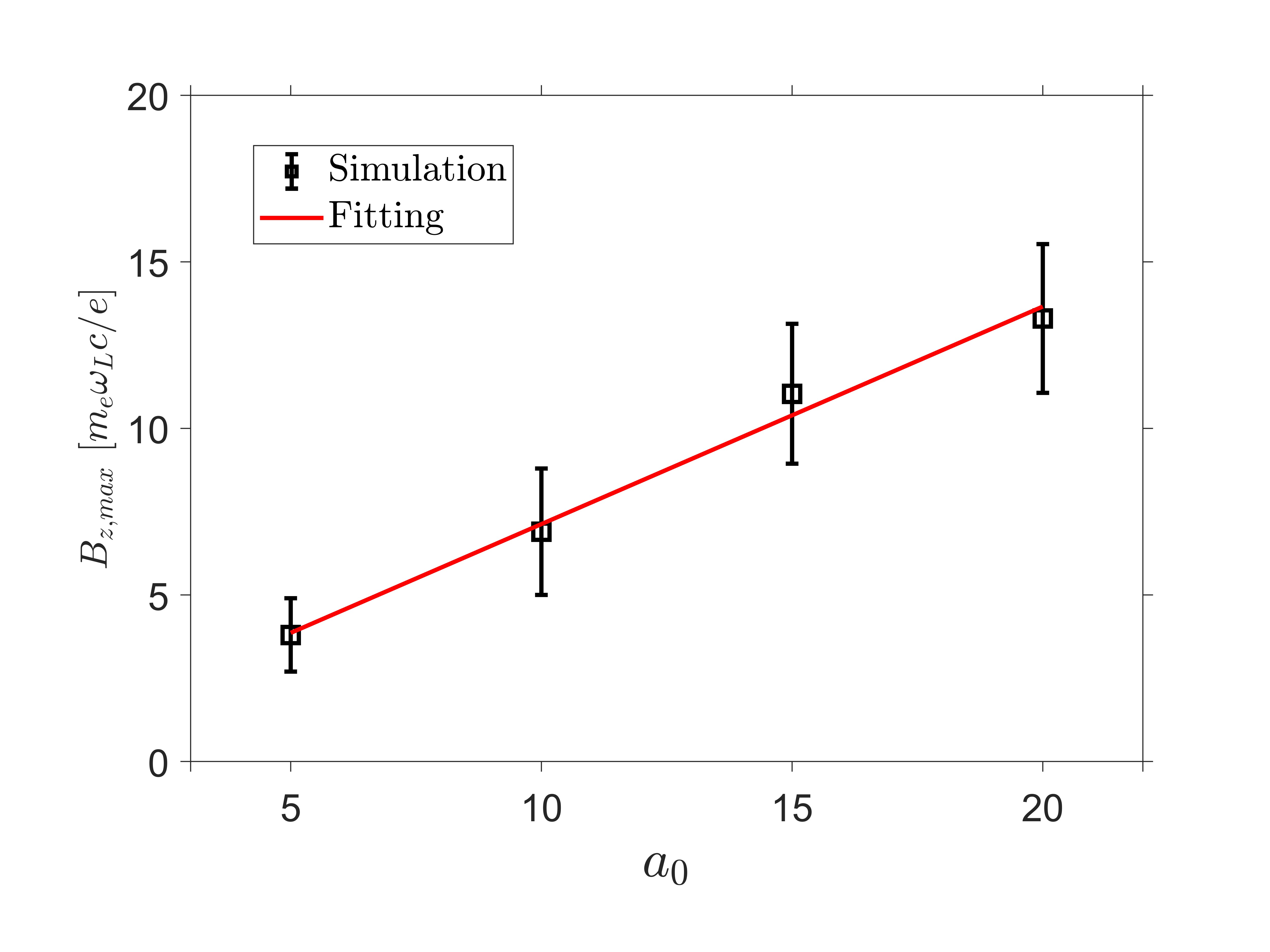}
\caption{Scaling law of the shear-flow magnetic field with the normalized laser amplitude. The red solid line represents the least-squares fit. Error bars for the magnetic field data are determined from the statistical variation of local maximum value within the shock region.}
\end{figure}

The scaling law relating the shear-flow magnetic field to the laser intensity is key to experimental design and can be derived from theoretical analysis and PIC simulations. The maximum field, given by Eq.\ 3, is
\begin{equation}
B_{z,max}=-\sqrt{\eta\bar{\gamma}}m_e\omega_{pe}u_0/e\propto n_e^{1/2}u_0\bar{\gamma}^{1/2},
\end{equation}
The maximum field depends on the electron density $n_e$, ion fluid velocity $u_0$, and the average electron Lorentz factor $\bar{\gamma}$ in the shock region. As $n_e$ is determined by the target design, it is incorporated into a scaling coefficient. The ion fluid velocity scales with the ion sound speed, $u_0\propto c_s=\sqrt{ZT_e/m_i}$, where the hot electron temperature $T_e$ (generated by the ultraintense laser) scales with the normalized amplitude $a_0$ according to the established scaling law \cite{Wilks}. This gives $u_0\propto a_0^{1/2}$. Similarly, $\bar{\gamma}\propto T_e\propto a_0$. Consequently, the scaling law of shear-flow magnetic field and laser intensity takes the form
\begin{equation}
B_{z,max}=\kappa a_0.
\end{equation}
The coefficient $\kappa$ depends on target parameters. For the target configuration used in this work, simulations performed at different laser intensities show that the shear-flow magnetic field exhibits a linear dependence on the laser intensity (Fig.\ 4). A least-squares fit yields a scaling coefficient of $\kappa \approx 0.7m_e\omega_L c/e$ ($\omega_L$ is laser angular frequency).

The shear-flow magnetic fields can reduce the time required for shock formation in ultraintense-laser-produced counter-streaming plasmas. The shock formation time $\tau$ is comparable to the ion gyro-period, 
\begin{equation}
\tau\sim 2\pi r_c/u_0=2\pi m_i c/ZeB_z.
\end{equation}
By substituting the simulated magnetic field ($B_z\sim \SI{200}{MG}$) from Fig.\ 2 into Eq.\ 6, the shock formation time is estimated to be $\tau = \SI{3}{ps}$, which closely matches the simulation results. For comparison, in high-power laser experiments conducted by Fiuza $et~al.$ at the NIF \cite{Fiuza}, shock formation requires several nanoseconds, nearly 3 orders of magnitude longer than that observed in our simulation. This ultrafast shock formation is attributed to the significant enhancement of magnetic field amplitude. In our simulations, all 3 parameters governing the magnetic field amplitude are significantly larger than those in the NIF experiments ($n_e\sim \SI{E22}{cm^{-3}}$, $u_0\sim \SI{3E4}{km/s}$ and $\bar{\gamma}\sim25$ for our simulation, $n_e\sim \SI{E20}{cm^{-3}}$, $u_0\sim \SI{2E3}{km/s}$ and $\bar{\gamma}\sim1$ for the NIF experiments).

Shear-flow magnetic fields enable collisionless shock formation at substantially reduced temporal and spatial scales, significantly lowering the required laser energy compared to high-power laser experiments. 
In the Appendix, we present a simulation (transverse width is \SI{50}{\mu m}) using a laser energy of $\sim \SI{10}{kJ}$, a factor of 50 lower than that used in the NIF experiments (\SI{0.455}{MJ}), in which an IW-CS still forms. The longitudinal scale of the counter-streaming plasmas required for shock formation is determined by ion fluid velocity and shock formation time in shock-region, $L\sim u_0\tau$. In our simulations, the ultrafast shock formation reduces this longitudinal scale by 2 orders of magnitude compared to the NIF experiments, which significantly increases the shock-region density (by 2 orders of magnitude). The transverse scale required for shock formation is characterized by the width of the shear-flow magnetic field, $w\sim 10\sqrt{\eta\bar{\gamma}}c/\omega_{pe}$ (estimated by the transverse width of \SI{50}{\mu m}), in our simulations. The increased shock-region density leads to a transverse scale approximately 2 orders of magnitude smaller than that in the NIF experiments ($w\sim 300 c/\omega_{pi}$ \cite{Kato2,Park}). The laser energy required for shock formation scales as $E_L\propto I\tau w^2$ (laser intensity $I\sim \SI{E20}{W/cm^2}$ for our simulation, $I\sim \SI{E15}{W/cm^2}$ for the NIF experiments), indicating that our simulations reduces the required laser energy by nearly 2 orders of magnitude compared to the NIF experiments. Facilities such as OMEGA-EP \cite{Kelly}, GEKKO-LFEX \cite{Miyanaga}, SG-II-UP \cite{Zhu}, and NIF-ARC \cite{Barty} already deliver multi-kilojoule, picosecond-scale laser pulses, making them suited for conducting IW-CSs experiments.

In summary, two-dimensional PIC simulations demonstrate that collisionless shocks in picosecond-scale ultraintense-laser-produced counter-streaming plasmas are mediated by shear-flow magnetic fields. We derive a one-dimensional analytical expression for the shear-flow magnetic fields and establish a scaling law relating their amplitude to laser intensity. Key findings reveal that such fields reduce shock formation time by 3 orders of magnitude and lower required laser energy by 2 orders of magnitude compared to the high-power laser experiments.

\begin{acknowledgments}
This work is supported by the National Science Foundation of China (No.\ 12147103).
\end{acknowledgments}


\begin{acknowledgments}
\end{acknowledgments}


\clearpage

$Appendix:~Simulation~with~reduced~transverse~scale$ - A simulation case with lower laser energy was carried out. This case reduced the simulation transverse scale to \SI{50}{\mu m}, other parameters are the same as the Fig.\ 2. The laser energy on each target is $E_L=I\tau\pi r^2$, where laser intensity $I=\SI{1.38E20}{W/cm^2}$, radius of laser focus spot is $r=\SI{25}{\mu m}$, laser duration is $\tau=\SI{4}{ps}$. So $E_L=\SI{10.8}{kJ}$. Fig.\ 5 shows the formation of both the shear-flow magnetic fields and the IW-CS, consistent with the results presented in Fig.\ 2.

\begin{figure}
\includegraphics[width=8cm]{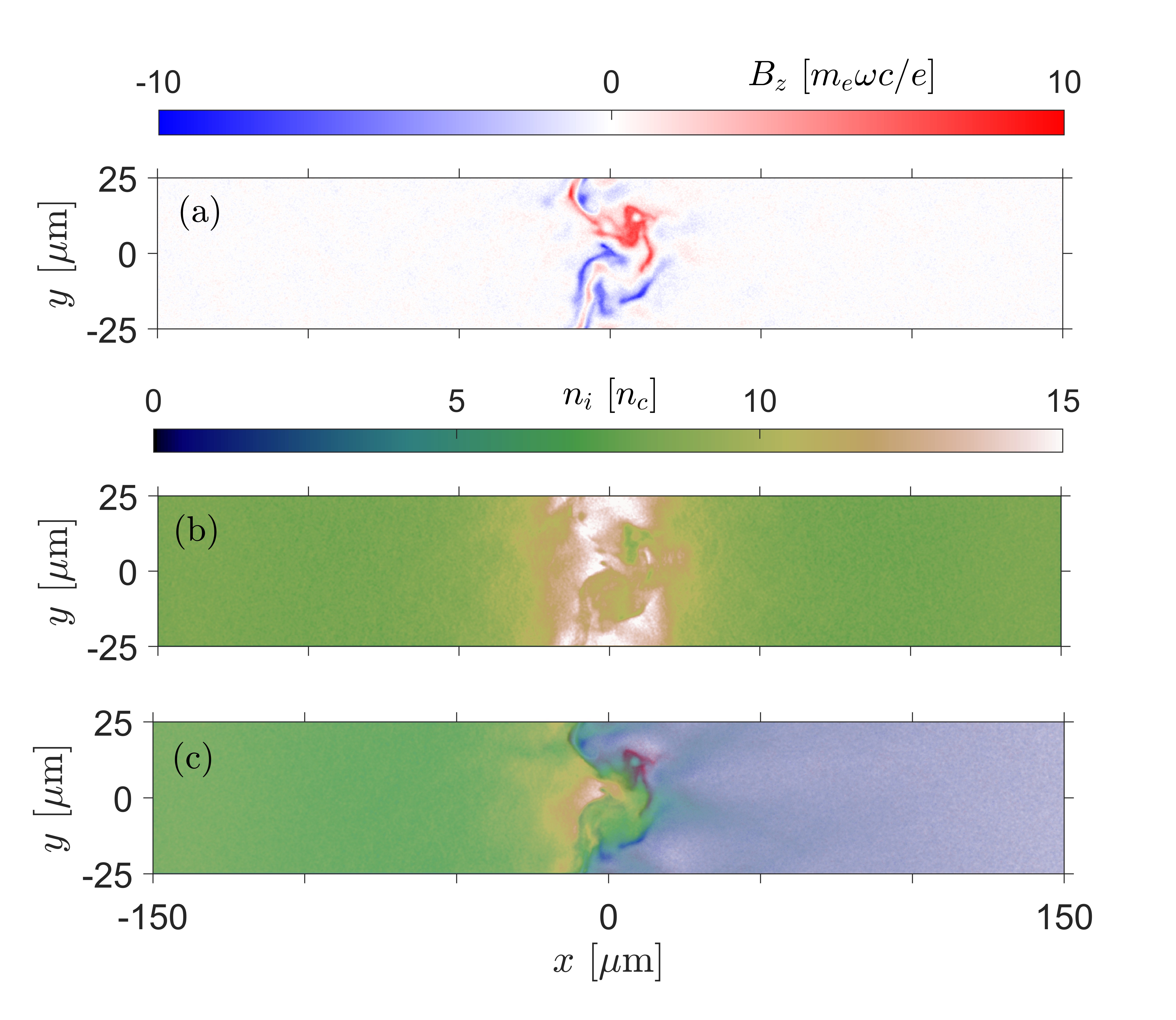}
\caption{Ultraintense-laser-produced collisionless shock in a reduced transverse scale simulation (at \SI{16}{ps}). (a) Self-generated magnetic field. (b) Ion density. (c) Magnetic field overlaid with ion density of left plasma flow.}
\end{figure}

\end{document}